# Crues exceptionnelles de la Vésubie et de la Roya (octobre 2020) : caractérisation hydrogéomorphologique et perspectives de gestion

## Major floods of the Vésubie and Roya Rivers (Alps, France) in October 2020: hydrogeomorphological caracterisation and management perspectives

Gabriel Melun[1], Frédéric Liébault[2], Guillaume Piton[2], Margot Chapuis[3], Paul Passy[4], Céline Martins[5] et Damien Kuss[5]

[1]Office français de la biodiversité (OFB) – gabriel.melun@ofb.gouv.fr
[2]Université Grenoble Alpes, INRAE, ETNA – frederic.liebault@inrae.fr ; guillaume.piton@inrae.fr
[3]Université Côte d'Azur, CNRS, ESPACE – margot.chapuis@unice.fr
[4]Université de Paris – paul.passy@u-paris.fr
[5]Office National des Forêts (ONF) – Services de Restauration de Terrains en Montagne (RTM) – celine.martins@onf.fr ; damien.kuss@onf.fr

## RÉSUMÉ

Le 2 octobre 2020, sous l'effet combiné de la tempête hivernale Alex formée au large de la Bretagne et d'un fort épisode méditerranéen, des pluies diluviennes ont affecté les bassins versants de la Roya et de la Vésubie (Alpes-Maritimes), avec des valeurs de précipitations en plusieurs endroits supérieures à 600 mm en 24h. Cet évènement paroxystique au lourd bilan humain (10 morts, 8 disparus), a engendré des crues éclair extrêmes sur une grande partie du réseau hydrographique de ces vallées alpines. Il en résulte une métamorphose fluviale quasi-généralisée des rivières, passant de tracés sinueux à chenal unique à des morphologies en tresses. La caractérisation des effets morphologiques de ces crues s'appuie notamment sur un travail de photo-interprétation diachronique qui permet de mettre en évidence une forte augmentation de la largeur de la bande active (jusqu'à 900 %) atteignant – voire repoussant sur certains secteurs – les limites antérieures du fond de vallée. Sur la Vésubie, l'effet morphologique 2D de la tempête Alex a été 10 fois supérieur à celui de la crue centennale de novembre 1997. Des travaux de comparaison de modèles numériques de terrain (MNT) avant/après crue permettent également d'entrevoir les variations altitudinales (érosion/dépôt) ayant affecté les lits et leurs marges. L'analyse des bouleversements entraînés par ces crues modifie la perception de l'espace de liberté de ces rivières torrentielles, qui doit être pris en compte dans la perspective d'une reconstruction résiliente.

## ABSTRACT

On October 2[nd], 2020, under the combined effect of the winter Alex storm formed off the Brittany coast, and a strong Mediterranean episode, very intensive rainfalls affected in the south eastern France, both Roya and Vésubie catchments (locally up to 600 mm in 24h). This paroxysmal event with a heavy human toll (10 dead, 8 missing) generated extreme flash floods over a large part of the hydrographic network. The result is an almost generalized fluvial metamorphosis of rivers, from sinuous single-thread channels to braided channels. The characterization of morphological effects of these floods is based on a diachronic aerial picture analysis highlighting a strong increase of the active channel width (up to 900%) reaching - or even pushing back in few sectors - front limits of the valley bottom. In the Vésubie, the 2D morphological effect of the Alex storm was 10 times higher than that of the 100-yrs return period flood of November 1997. Comparison of digital terrain models (DEM) before- and after-flood also allows us to foresee the altitudinal variations (erosion/deposition) that affected beds and their riverine margins. The analysis of the impacts caused by these floods changes the perception of the "freedom space" of these alpine rivers, which now must be taken into account in the perspective of resilient reconstruction.

## MOTS CLÉS

Analyse diachronique, Crue torrentielle, Métamorphose fluviale, Photo-interprétation, Roya, Vésubie

# 1 INTRODUCTION

## 1.1 Des précipitations à la réponse hydrologique

Le 2 octobre 2020 des précipitations exceptionnelles se sont abattues sur plusieurs vallées maralpines et tout particulièrement sur celles de la Vésubie et de la Roya, où les cumuls enregistrés sur 24h ont localement pu dépasser 650 mm. Ces précipitations ont engendré une rapide montée des eaux dans les thalwegs ; les niveaux d'eau des principales rivières gagnant 6 à 8 m en quelques heures. Les débits dépassent toutes les références connues : la Vésubie atteint un débit d'environ 900 m$^3$/s à l'exutoire et la Roya a probablement dépassé 1200 m$^3$/s pour des rivières dont le module est inférieur à 10 m$^3$/s (CEREMA, 2021). Dans la plupart des cas, les débits centennaux estimés, sont largement dépassés.

## 1.2 Un bouleversement géomorphologique

Les vallées de la Vésubie et de la Roya présentent des prédispositions géomorphologiques à la formation de vagues sédimentaires (fortes pentes, potentiel de recharge sédimentaire important depuis les têtes de bassin, les lits majeurs et les terrasses alluviales). Les crues majeures d'octobre 2020 ont ainsi été marquées par une importante métamorphose fluviale des cours de la Vésubie et de la Roya ; métamorphose caractérisée par l'émergence quasi-systématique de vastes bandes de tressage, associées à un engravement pluri-métrique des lits (Fig.1).

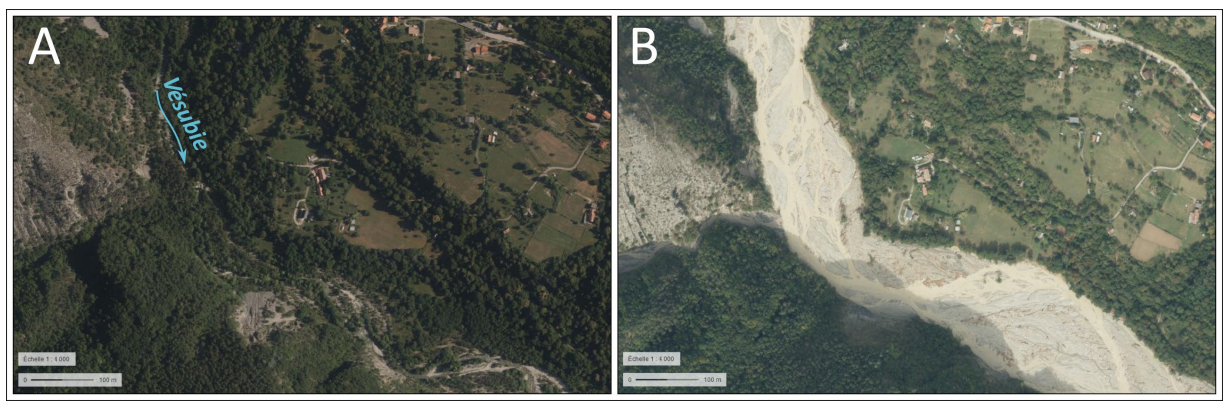

Figure1 : Métamorphose fluviale de la Vésubie en aval de Saint-Martin Vésubie suite à la crue du 2 octobre 2020 (source : orthophotos IGN 2017 (A) et 2020 post-crue (B)).

# 2 MÉTHODES

## 2.1 Caractérisation 2D des changements morphologiques

La caractérisation précise des impacts hydrogéomorphologiques des crues d'octobre 2020 s'appuie sur la photo-interprétation diachronique de 7 couvertures d'orthophotographies (IGN et CRIGE-PACA) couvrant une période allant de 1948 à 2020 (post-crue). Les fonds de vallées modernes (dépôts fluviatiles historiques) et les bandes actives de la Vésubie, de la Roya et de certains de leurs principaux affluents ont été digitalisés. Une procédure de segmentation spatiale à partir d'un référentiel commun (fond de vallée post-Alex) a été réalisée sous QGIS de façon à extraire les largeurs de bande active et du fond de vallée et à analyser le gradient longitudinal de la reconquête de bande active.

## 2.2 Analyses 3D des fonds de vallées

Une analyse comparative des modèles numériques de terrain (MNT) LiDAR avant-après crue, mis à disposition par l'IGN et la Métropole Nice Côte d'Azur est également proposée afin d'appréhender les variations altitudinales (exhaussement, incision des lits, érosion latérale) ayant affecté les vallées.

# 3 RÉSULTATS & DISCUSSIONS

L'analyse historique des photographies aériennes montre que la Vésubie et la Roya présentent des bandes actives déjà rétractées dès la première moitié du XX$^e$ siècle, différant ainsi d'autres vallées alpines (Lallias Tacon *et al*., 2017). Les différentes crues récentes (1997, 2020) sont bien à l'origine d'une forte respiration du lit, traduite par un accroissement de la largeur moyenne des bandes actives et une aggradation du fond du lit. Les crues d'octobre 2020 sont dans ce sens particulièrement remarquables : elles ont entraîné une augmentation forte et généralisée de la largeur moyenne des lits de la Vésubie, de la Roya et de leurs principaux affluents (Fig. 2A). Cette augmentation est particulièrement significative (facteur 4) dans les secteurs où les fonds de vallée sont plus larges, et non contraints pas des secteurs en gorges. Localement, les largeurs de bandes actives ont pu être multipliées par 10, passant de chenaux sinueux unique de 15 m de largeur, à une bande de tressage d'une largeur supérieure à 150 m. Sur certains secteurs ce sont les limites mêmes du fond de vallée qui ont été repoussées par les écoulements de crues, modifiant ainsi l'espace maximal potentiel de divagation latéral (Fig. 2B).

Les largeurs atteintes par les bandes actives post-crue permettent de rapprocher la Vésubie du modèle régional établi pour les rivières alpines en tresses (Piégay *et al.,* 2009). Ceci ressort particulièrement dans les parties amont du bassin versant, qui ont été plus fortement impactées par les vagues sédimentaires ; tout comme certains secteurs bien localisés de bassin de la Roya (Madone de Viévola, amont de Tende, aval du torrent de la Bieugne). Ces résultats témoignent d'une manière générale de l'intensité remarquable de la fourniture et du transport sédimentaire lors de la crue.

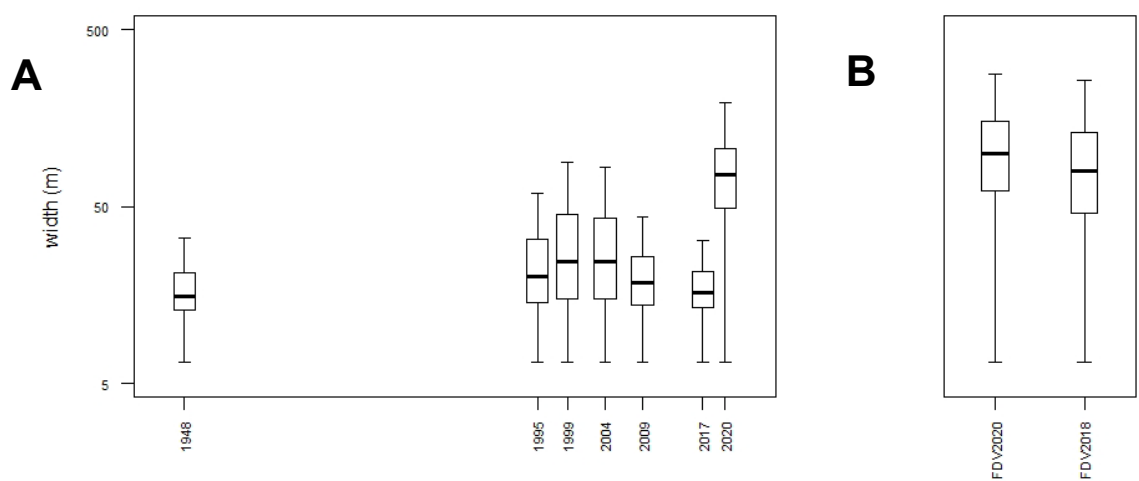

Figure 2 : Évolution diachronique de la largeur moyenne du lit de la Vésubie entre 1948 et 2020 (A), et du fond de vallée entre 2018 et 2020 (B).

L'analyse comparative des MNT est en cours. Les pré-analyses mettent en évidence de fortes variations (métrique à décamétrique) de l'altitude des lits et de leurs marges. Des processus très significatifs d'érosion des berges et pieds de versant ayant localement pu dépasser 1000 m$^3$/ml semblent avoir fourni les importantes quantités de matériaux qui ont comblé partiellement les fonds de vallées. Des résultats plus aboutis seront présentés lors de la conférence.

## 4 CONCLUSION

Les analyses réalisées permettent de mettre en évidence des évolutions géomorphologiques sans précédents sur l'ensemble des vallées de la Vésubie et de la Roya depuis plus d'un siècle. Les résultats de ces travaux de caractérisation hydrogéomorphologique permettent de rapprocher les crues d'octobre 2020 de certaines reconquêtes de bande active post-crues exceptionnelles observées au XX$^e$ siècle en France, à l'image de celles de la Tech et du Têt en 1940 ou du Guil en 1957.

## BIBLIOGRAPHIE

CEREMA. 2021. RETEX technique ALEX - Inondations des 2 et 3 octobre 2020 - Consensus hydrologique. Rapport établi pour le compte de la DDTM06, version 2 du 14 Sept., 59 p.

Lallias-Tacon, S., Liébault, F. et Piégay, H. (2017). Use of airborne LiDAR and historical aerial photos for characterising the history of braided river floodplain morphology and vegetation responses. *CATENA*, 149, Part 3: 742-759.

Piégay H, Alber A, Slater L, Bourdin L. (2009). Census and typology of braided rivers in the French Alps. *Aquatic Sciences,* 71: 371-388.